\documentclass[aps,showpacs,superscriptaddress,floatfix]{revtex4}
\usepackage{amssymb,amsmath}
\usepackage[dvips]{graphicx,color}

\def\re{\mbox{Re}\,}
\def\openone{\leavevmode\hbox{\small1\kern-3.8pt\normalsize1}}%

\def\mf{{\mbox{\tiny\em MFA}}}

\def\bea{\begin{eqnarray}}
\def\eea{\end{eqnarray}}
\def\beq{\begin{equation}}
\def\eeq{\end{equation}}

\begin{document}

\title{Phase diagram of neutral quark matter in
nonlocal chiral quark models}

\author{D. G\'omez Dumm}
\email{dumm@fisica.unlp.edu.ar}
\affiliation{IFLP, CONICET $-$ Dpto.\ de F\'{\i}sica,
     Universidad Nacional de La Plata,
     C.C.\ 67, 1900 La Plata, Argentina}
\affiliation{CONICET, Rivadavia 1917, 1033 Buenos Aires, Argentina}

\author{D.B. Blaschke}
\email{blaschke@theory.gsi.de}
\thanks{present address: Institute of Physics, University of Rostock,
D-18055 Rostock, Germany}
\affiliation{Gesellschaft f\"ur Schwerionenforschung (GSI),
Planckstr.\ 1, 64291 Darmstadt, Germany \\
Bogoliubov  Laboratory of Theoretical Physics, JINR Dubna,
Joliot-Curie Street 6, 141980  Dubna, Russia}

\author{A.G. Grunfeld}
\email{grunfeld@tandar.cnea.gov.ar}
\affiliation{Physics Department, Comisi\'on Nacional de Energ\'{\i}a At\'omica,
 Av.\ Libertador 8250, 1429 Buenos Aires, Argentina}

\author{N.N.\ Scoccola}
\email{scoccola@tandar.cnea.gov.ar}
\affiliation{CONICET, Rivadavia 1917, 1033 Buenos Aires, Argentina}
\affiliation{Physics Department, Comisi\'on Nacional de Energ\'{\i}a At\'omica,
 Av.\ Libertador 8250, 1429 Buenos Aires, Argentina}
\affiliation{Universidad Favaloro, Sol{\'\i}s 453, 1078 Buenos Aires,
Argentina}

\begin{abstract}
We consider the phase diagram of two-flavor quark matter under neutron
star constraints for two nonlocal, covariant quark models within the mean
field approximation. In the first case (Model I) the nonlocality arises
from the regularization procedure, motivated by the instanton liquid
model, whereas in the second one (Model II) a separable approximation of
the one-gluon exchange interaction is applied. We find that Model II
predicts a larger quark mass gap and a chiral symmetry breaking (CSB)
phase transition line which extends 15-20 \% further into the phase
diagram spanned by temperature ($T$) and chemical potential ($\mu$). The
corresponding critical temperature at $\mu=0$, $T_c(0) \simeq 140$ MeV, is
in better accordance to recent lattice QCD results than the prediction of
the standard local NJL model, which exceeds 200 MeV. For both Model I and
Model II we have considered various coupling strengths in the scalar
diquark channel, showing that different low-temperature quark matter
phases can occur at intermediate densities: a normal quark matter (NQM)
phase, a two-flavor superconducting (2SC) quark matter phase and a mixed
2SC-NQM phase. Although in most cases there is also a gapless 2SC phase,
this occurs in general in a small region at nonzero temperatures, thus its
effect should be negligible for compact star applications.
\end{abstract}

\date{\today}


\maketitle

\renewcommand{\thefootnote}{\arabic{footnote}}
\setcounter{footnote}{0}

\section{Introduction}

Within the last decade the investigation of the thermodynamics and phase
structure of strongly interacting matter has been driven by the results of
the experimental programs with ultrarelativistic heavy ion beams at
CERN-SPS and BNL-RHIC \cite{Ritter:2004xj} as well as by the unprecedented
quality of data from lattice QCD simulations \cite{Karsch:2003vd}. A new
picture of the state of matter created in these experiments has emerged,
according to which the physical nature of the sought-for quark-gluon
plasma (QGP) is a perfect liquid of strongly correlated hadron-like
resonances rather than an ideal gas of quasifree quarks and gluons
\cite{Shuryak:2004kh,Blaschke:2003ut}. The persistence of a strong,
nonperturbative coupling in the QGP can be supposed as a prerequisite of a
successful modelling of the QCD phase diagram. Turning to the domain of
finite chemical potentials, guidance from lattice QCD is limited to $\mu
\le T$ (where Taylor expansion techniques can be applied), and
experimental programmes such as CBM at FAIR Darmstadt are yet in the stage
of preparation with a planned begin of operation in 2014. Therefore,
predictions for the QCD thermodynamics at low temperatures and high baryon
densities $\mu \ge T$, where the critical point and the regions of color
superconducting quark matter are expected in
the QCD phase diagram~\cite{Rev}, have to be developed within effective
models for nonperturbative QCD and to be tested against observational
constraints from neutron stars \cite{Blaschke:2001uj}. For a recently
developed testing scheme we refer to \cite{Klahn:2006ir}.

After the early discussion on color superconducting dense quark
matter~\cite{Bailin:1983bm} was revived for nonperturbatively strong
couplings within effective quark models~\cite{ARW98}, a great variety of
possible diquark pairing patterns and corresponding phases has been
explored~\cite{Rev} and estimates for the order of magnitude of the pairing
gaps have been given. One of the central questions for phenomenological
applications in compact stars, where electric and charge neutrality has to be
imposed~\cite{Iida:2000ha,AR02}, concerns the number of active flavors. It
turns out that at low temperatures one obtains a sequential melting pattern
of the light and strange quark chiral condensates, which is rather
insensitive to the details of the four-momentum dependence of the
interaction, but crucially dependent on whether also the strange quark mass
is determined selfconsistently (see~\cite{Gocke:2001ri} for an early work
within the covariant chiral quark model). The present 'state-of-the-art' for
the description of color superconducting phases is represented by the fully
selfconsistent mean field three-flavor Nambu-Jona-Lasinio (NJL) model
calculations of Refs.~\cite{Ru05,Bla05,Abuki:2005ms}. The results of these
studies indicate that under compact star conditions, i.e.\ color and electric
charge neutrality together with $\beta-$ equilibrium, the two-flavor color
superconducting (2SC) phase is favored over the three-flavor
color-flavor-locking (CFL) one. This outdates earlier results along the lines
of, e.g., Ref. \cite{AR02} where the strange quark mass has been set to a
constant, small value. Even if the third quark flavor occurs at not too high
densities to be in principle realized in compact star interiors, the star
configurations with CFL quark cores turn out to be hydrodynamically unstable
\cite{Baldo:2002ju}. For a more detailed introduction to the recent status of
dense quark matter in the NJL model, see~\cite{Bub04}.

Studies of neutral 2SC quark matter consider also the presence of a
so-called gapless (g2SC) phase~\cite{SH03}, which is found to occur at
intermediate temperatures and chemical potentials. Going beyond the NJL
theory, results arising from a noncovariant nonlocal quark
model~\cite{Blaschke:2003yn,Grigorian:2003vi} with Gaussian regulator
functions show that the 2SC phase is not present in asymmetric quark
matter for standard values of the diquark coupling~\cite{ABG05}. In this
model, however, for strong diquark couplings one does find a 2SC phase,
together with a g2SC and a mixed normal quark matter (NQM)-2SC phase in
which the electric charge neutrality is satisfied only
globally~\cite{Bed02,Shovkovy:2003ce,Neumann:2002jm}.

As the structure of the quark matter phase diagram within NJL-type models is
settling, it is important to investigate the situation in the case of
effective models that go beyond those used in
Refs.~\cite{AR02,Ru05,Bla05,Abuki:2005ms,ABG05}, in the sense that they
include fully covariant nonlocal interactions. Nonlocality arises naturally
in the context of several quite successful approaches to low energy quark
dynamics as, for example, the instanton liquid model~\cite{SS98} and the
Schwinger-Dyson resummation techniques~\cite{RW94}. The same happens in
lattice QCD~\cite{SLW01}. It has been also argued that nonlocal models have
several advantages over the local ones (e.g., the NJL model~\cite{NJL} and
its generalizations). Indeed, nonlocal interactions regularize the model in
such a way that anomalies are preserved~\cite{AS99} and charges properly
quantized, the effective interaction is finite to all orders in the loop
expansion and there is no need to introduce extra cut-offs. Soft regulators
such as Gaussian functions lead to small next-to-leading order
corrections~\cite{Blaschke:1995gr,Rip00}, etc. This type of models has been
successfully used to investigate
meson~\cite{Ito:1991sz,Buballa:1992sz,BB95,SGS04} and
baryon~\cite{BGR02,RWB03,RP05} properties at vanishing temperature and
chemical potential. The phase diagram of isospin symmetric matter has also
been studied within this
context~\cite{Blaschke:1999ab,GGS01,DS05,DGS04,Blaschke:2004cc}. The aim of
the present work is to extend these analyses to the case in which compact
star conditions are imposed. {We will consider two-flavor versions of the
models, since (as mentioned below) effects arising from the strange quark can
be safely neglected in the region of intermediate chemical potentials to be
covered}. The investigation of the question whether the results of the above
referenced studies will remain qualitatively unchanged after the inclusion of
correlations beyond the mean-field level goes beyond the scope of the present
work and deserves a separate study. It should be noticed, however, that in
the non-strange case considered here the relevant color superconducting
condensate transforms as a singlet representation of the global
SU(2)$_L\;\otimes\;$SU(2)$_R$ chiral group. Thus, no global symmetry is
broken by the 2SC ground state, which implies that there are no Goldstone
bosons that could further condense.

The paper is organized as follows. In Section II we introduce the model and
derive some useful analytical expressions. In Section III we present the
numerical results obtained for the case of a Gaussian regulator, considering
different ratios between the coupling constants. Then, in Section IV we
discuss the features of the obtained phase diagrams, and in Section V we
present our conclusions.

\section{Formalism}

Let us begin by stating the Euclidean action for the nonlocal chiral quark
model in the case of two light flavors and anti-triplet diquark
interactions,
\begin{equation}
S_E = \int d^4 x \ \left\{ \bar \psi (x) \left(- i \rlap/\partial
+ m \right) \psi (x) -
\frac{G}{2} j^f_M(x) j^f_M(x) -
\frac{H}{2} \left[j^a_D(x)\right]^\dagger j^a_D(x) \right\} \,.
\label{action}
\end{equation}
Here $m$ is the current quark mass, which is assumed to be equal for $u$
and $d$ quarks. The nonlocality can be introduced now in different
ways~\cite{Rip97}. In what follows we will work within two alternative
scenarios, that we call ``Model I'' and ``Model II'', in which the mesonic
current $j_{M}(x)$ and the diquark current $j_{D}(x)$ in Eq.~(\ref{action})
are given by nonlocal operators.
In the case of Model I~\cite{Buballa:1992sz,BB95}, the effective
interactions are based on an instanton liquid picture of QCD.
The nonlocal currents read
\begin{eqnarray}
j^f_M (x) &=& \int d^4 y\ d^4 z \ r(y-x) \ r(x-z) \ \bar \psi(y)
\ \Gamma_f \ \psi(z)\,, \nonumber \\
j^a_D (x) &=& \int d^4 y\ d^4 z \ r(y-x) \ r(x-z) \ \bar \psi_C(y) \
i \gamma_5 \tau_2 \lambda_a \ \psi(z)\,,
\label{uno}
\end{eqnarray}
where we have defined $\psi_C(x) = \gamma_2\gamma_4 \,\bar \psi^T(x)$ and
$\Gamma_f = ( \openone, i \gamma_5 \vec \tau )$, while $\vec \tau$ and
$\lambda_a$, with $a=2,5,7$, stand for Pauli and Gell-Mann matrices acting
on flavor and color spaces, respectively.

On the other hand, Model II~\cite{Ito:1991sz,SBK94} arises from a
separable form of the effective one-gluon exchange (OGE) picture. In this
case the nonlocal currents $j_{M,D}(x)$ are given by
\begin{eqnarray}
j^f_M (x) &=& \int d^4 z \  g(z) \
\bar \psi(x+\frac{z}{2}) \ \Gamma_f \ \psi(x-\frac{z}{2})\,,
\nonumber \\
j^a_D (x) &=&  \int d^4 z \ g(z)\ \bar \psi_C(x+\frac{z}{2}) \
i \gamma_5 \tau_2 \lambda_a \ \psi(x-\frac{z}{2}) \; .
\label{cuOGE}
\end{eqnarray}
The functions $r(x-y)$ and $g(z)$ in Eqs.~(\ref{uno}) and (\ref{cuOGE})
are nonlocal regulators characterizing the corresponding interactions. It
is convenient to translate them into momentum space. Since Lorentz
invariance implies that they can only be functions of $p^2$, we will use
for the Fourier transforms of these regulators the forms $r(p^2)$ and
$g(p^2)$ from now on.

The current-current interaction in Eq.~(\ref{action}) is quite common to
effective theories for low-energy QCD such as the NJL model. The momentum
dependence introduced by the functions $r(p^2)$ and $g(p^2)$ is a
generalization of that model, and can be chosen in such a way that the
dynamical mass function of the quark propagator reproduces some features
obtained in lattice QCD analyses. It could entail, e.g., the possibility
of complex conjugate mass poles of the quark propagator, in agreement with
results arising from the Schwinger-Dyson equation approach to QCD. The
index structure of the vertices $\Gamma_f$ and also the ratio of the
coupling constants $H/G$ can be obtained from a Fierz rearrangement of
e.g.\ the OGE interaction, as detailed in \cite{Bub04} (see also
references therein). For the OGE, as well as for the instanton liquid
model, one obtains $H/G=0.75$. However, since a precise derivation of the
effective couplings from QCD is not known, there is a significant
theoretical uncertainty in this value. In fact, so far there is no strong
phenomenological constraint on $H/G$, except for the fact that values
larger that $H/G \sim 1$ are quite unlikely to be realized in QCD, since
they might lead to color symmetry breaking in the vacuum. We will leave
this ratio as a free parameter, analyzing the results obtained for values
lying within a range from 0.5 to 1.

The partition function of the system at temperature $T$ and quark chemical
potentials $\mu_{fc}$ is given by
\begin{equation}
{\cal Z} = \int {\cal D} \bar\psi\, {\cal D} \psi \
e^{-S_E(\mu_{fc},T)} \ ,
\label{zcero}
\end{equation}
where the Euclidean action is obtained from Eq.~(\ref{action}) by going to
momentum space and performing the replacements
\begin{equation}
p_4 \ \ \to \ \ \omega_n - i\mu_{fc}\ \ , \qquad\qquad
\int \frac{d^4 p}{(2\pi)^4} \ \ \to \ \ T
\sum_{n=-\infty}^{\infty} \int \frac{d^3 \vec{p}}{(2\pi)^3}\ .
\end{equation}
Here $p_4$ is the fourth component of the (Euclidean) momentum of a quark
carrying flavor $f$ and color $c$, and $\omega_n$ are the Matsubara
frequencies corresponding to fermionic modes, $\omega_n = (2 n+1) \pi T$.
We are assuming here that quark interactions depend on the temperature and
chemical potentials only through the arguments of the regulators. Note
that, as required for the applications we are interested in, we have
introduced in Eq.~(\ref{zcero}) different chemical potentials for each
quark flavor and color.

To proceed it is convenient to perform a standard bosonization of the theory.
This procedure is described, e.g., in Refs.~\cite{Rip97,SBK94}. Thus, we
introduce the bosonic fields $\sigma$, $\pi_a$ and $\Delta_a$, and integrate
out the quark fields. In what follows, we work within the mean field
approximation (MFA), in which these bosonic fields are replaced by their
vacuum expectation values $\bar \pi_a=0$, $\bar \sigma$ and $\bar \Delta_a$.
Moreover, we adopt the usual 2SC ansatz $\bar \Delta_5= \bar \Delta_7=0$,
$\bar \Delta_2= \bar \Delta$. We have assumed other possible condensates
to be negligible, following previous analyses carried out in the NJL model
framework~\cite{Rev,Bub04}. Within the MFA, and employing the Nambu-Gorkov
formalism, the thermodynamical potential per unit volume can be
written as
\begin{equation}
\Omega^\mf  =  - \frac{T}{V} \, \ln {\cal
Z}^\mf  = \frac{ \bar \sigma^2 }{2 G} + \frac{|\bar
\Delta|^2}{2 H} - \frac{T}{2} \sum_{n=-\infty}^{\infty} \int \frac{d^3
\vec{p}}{(2\pi)^3} \
\ln \mbox{det} \left[ \frac{1}{T} \ S^{-1}(\bar \sigma ,\bar \Delta)
\right] \ . \label{z}
\end{equation}
Here the inverse propagator $S^{-1}(\bar \sigma,\bar \Delta)$ is a $48
\times 48$ matrix in Dirac, flavor, color and Nambu-Gorkov spaces, given
by
\begin{equation}
\left(%
\begin{array}{cccccccc}
  -\rlap/p_{ur}^+  + \Sigma_{ur} & 0 & 0 & 0
               & 0 & 0 & \gamma_5 \tau^c_2 \ \Delta & 0
\\
  0 &  -\rlap/p_{ub}^+  + \Sigma_{ub} & 0 & 0
               & 0 & 0 & 0 & 0
\\
  0 & 0 & -\rlap/p_{dr}^+  + \Sigma_{dr} & 0
               & -\gamma_5 \tau^c_2 \  {\Delta}^* & 0 & 0 & 0
\\
  0 & 0 & 0 & -\rlap/p_{db}^+  + \Sigma_{db}
               & 0 & 0 & 0 & 0
\\
  0 & 0 & \gamma_5 \tau^c_2 {\Delta}^* & 0
               & -\rlap/p_{ur}^-  + {\Sigma_{ur}}^* & 0 & 0 & 0
\\
  0 & 0 & 0 & 0
               & 0 & -\rlap/p_{ub}^-  + {\Sigma_{ub}}^* & 0 & 0
\\
  -\gamma_5 \tau^c_2 \Delta  & 0 & 0 & 0
               & 0 & 0 & - \rlap/p_{dr}^-  +  {\Sigma_{dr}}^* & 0
\\
  0 & 0 & 0 & 0
               & 0 & 0 & 0 & -\rlap/p_{db}^-  + {\Sigma_{db}}^* \\
\end{array}%
\right)
\label{matrix}
\end{equation}
where we have used the definitions
\begin{eqnarray}
p_{fc}^{\pm} & = & \left(\; \omega_n \mp i \,\mu_{fc}
\ , \ \vec p\ \right) \ , \\
\Sigma_{fc} & = & m \; + \; \bar \sigma \ h(p^+_{fc},p^+_{fc}) \ , \\
\Delta & = & \bar \Delta \ h(p^+_{ur},p^-_{dr}) \ , \label{deltah}
\end{eqnarray}
with $f = u,d$ and $c=r,g,b$. The functions $h(p,q)$ have been introduced
in order to have a common notation for both Model I and Model II. One has
\begin{eqnarray}
h(s,t)=\left\{\begin{array}{c}
r(s^2)\ r(t^2) \ \ \qquad \qquad \ \ {\rm (Model\ I)} \\
\\
g\Big(\Big[\frac{s+t}{2}\Big]^2\Big) \ \qquad \qquad \ \ {\rm (Model\ II)} \\
\end{array}
\right.
\label{hache}
\end{eqnarray}
We have taken into account that, as we will see below, the usual 2SC
ansatz implies $\mu_{fr} = \mu_{fg}$. In Eq.~(\ref{matrix}), entries with
subindices $ur$ and $dr$ are intended to be multiplied by an
$\openone_{2\times 2}$ matrix in $rg$ space, while $\tau^c_2$ stands for a
$\tau_2$ Pauli matrix acting in this space.

The determinant of $S^{-1}$ can be analytically evaluated, leading to
\begin{eqnarray}
\Omega^\mf &=& \frac{\bar \sigma^2}{2 G}\ +\
\frac{\bar \Delta^2}{2 H}\ -\ T \sum_{n=-\infty}^{\infty} \int \frac{d^3
\vec{p}}{(2\pi)^3} \sum_{c\,=\,r,g,b} \ln \frac{\;|A_c|^2 }{T^8} \ ,
\label{once}
\end{eqnarray}
where
\begin{equation}
A_c \ = \ ({p_{uc}^+}^2 + \Sigma_{uc}^2)\; \left({p_{dc}^-}^2 +
{\Sigma_{dc}^*}^2\right) \; + \;(1-\delta_{bc})\; {\Delta}\,^2\;
[{\Delta}\,^2 + 2\,(p_{uc}^+ \cdot p_{dc}^-) + 2\, \Sigma_{uc}
\,\Sigma_{dc}^*] \ . \label{def1}
\end{equation}

For finite values of the current quark mass, $\Omega^\mf$ turns out to be
divergent. The regularization procedure used here amounts to define
\begin{equation}
\Omega^\mf_{\rm (reg)} =
\Omega^\mf \ - \ \Omega^{\rm free} \ + \ \Omega^{\rm free}_{\rm (reg)}\ ,
\label{omegareg}
\end{equation}
where $\Omega^{\rm free}$ is obtained from Eq.~(\ref{once}) by setting
$\bar \Delta = \bar \sigma=0$, and $\Omega^{\rm free}_{\rm (reg)}$ is the
regularized expression for the thermodynamical potential of a free fermion
gas,
\begin{equation}
\Omega^{\rm free}_{\rm (reg)} = -2\ T \int \frac{d^3 \vec{p}}{(2\pi)^3}\; \sum_{f,c}
\left\{ \ln\left[ 1 + e^{-\left( \sqrt{\vec{p}^2+m^2}-\mu_{fc}
\right)/T} \right] + \ln\left[ 1 + e^{-\left(
\sqrt{\vec{p}^2+m^2}+\mu_{fc} \right)/T} \right] \right\}\ .
\label{freeomegareg}
\end{equation}

The mean field values $\bar \sigma$ and $\bar \Delta$ are obtained from
the coupled gap equations
\begin{eqnarray}
& & \frac{ d \Omega^\mf_{\rm (reg)}}{d\bar \Delta} \ = \
\bar \Delta\; ( 1 - 16\ H\ T \ D_{ud} ) \ = \ 0 \ , \label{deltu} \\
& & \rule{0cm}{.7cm} \frac{ d \Omega^\mf_{\rm (reg)}}{d\bar \sigma} \ = \
\bar \sigma - 4\; G\; T \ ( S_{ud} + S_{du} ) \ = \ 0 \ , \label{sigud}
\end{eqnarray}
where we have defined
\begin{eqnarray}
& & D_{ij} \ = \ D_{ji} \ = \ \re \!\sum_{n=-\infty}^{\infty} \int
\frac{d^3 \vec{p}}{(2\pi)^3} \ \ h^2(p^+_{ur},p^-_{dr})\ \frac{
\Delta^2 + (p_{ir}^+ \cdot p_{jr}^-) +
\Sigma_{ir}\,\Sigma_{jr}^*}
{A_r} \ , \label{dud} \\
& & S_{ij} \ = \ \re\!\! \sum_{n=-\infty}^{\infty} \!\int\!
\frac{d^3 \vec{p}}{(2\pi)^3} \left\{ 2\, h(p^+_{ir},p^+_{ir}) \ \frac{
\, \Sigma_{ir}\,({p_{jr}^-}^2 + {\Sigma^*_{jr}}^2) +
{\Delta}\,^2\,\Sigma^*_{jr}}{A_r} \ + \ h(p^+_{ib},p^+_{ib}) \
\frac{\Sigma_{ib}} {{p_{ib}^+}^2 + \Sigma_{ib}^2}\! \right\} \ \ .
\end{eqnarray}

So far we have introduced different chemical potentials for each quark
flavor and color. However, not all of them are in general independent
quantities. For the description of quark matter in the core of neutron
stars, we require the system to be color and electric charge neutral (for
a further discussion on the issue of color neutrality and color
singletness we refer to Refs.~\cite{AR02,Amore:2001uf}). Thus, within the
previously introduced 2SC ansatz, only one color-dependent chemical
potential is needed~\cite{BuSh05}, and the $\mu_{fc}$ can be written in
terms of only three independent quantities: the baryonic chemical
potential $\mu_B$, the quark electric chemical potential $\mu_{Q_q}$ and
the color chemical potential $\mu_8$. Defining $\mu\equiv\mu_B/3$, the
corresponding relations read
\begin{eqnarray}
\mu_{ur} = \mu_{ug} &=& \mu + \frac23 \mu_{Q_q} + \frac13 \mu_8 \nonumber \\
\mu_{dr} = \mu_{dg} &=& \mu - \frac13 \mu_{Q_q} + \frac13 \mu_8 \nonumber \\
\mu_{ub} &=& \mu + \frac23 \mu_{Q_q} - \frac23 \mu_8 \nonumber \\
\mu_{db} &=& \mu - \frac13 \mu_{Q_q} - \frac23 \mu_8
\label{chemical}
\end{eqnarray}

Now, in the core of neutron stars, in addition to quark matter we have
electrons. Thus, within the mean field approximation for the quark matter,
and considering the electrons as a free Dirac gas, the full grand
canonical potential is given by
\begin{equation}
\Omega^{full} = \Omega^\mf_{\rm (reg)} + \Omega^{e} \ ,
\end{equation}
where
\begin{equation}
\Omega^{e} = - \frac{1}{12 \pi^2} \left( \mu_e^4 + 2 \pi^2 T^2
\mu_e^2 + \frac{7\pi^4}{15} T^4 \right)\ ,
\end{equation}
$\mu_e$ being the electron chemical potential. For simplicity we have
neglected here the electron mass.

In addition, it is necessary to take into account that quark matter has to
be in $\beta-$ equilibrium with electrons through the $beta-$ decay reaction
\begin{equation}
d \rightarrow u + e + \bar \nu_e \ .
\end{equation}
Thus, assuming that antineutrinos escape from the stellar core, we
must have
\begin{equation}
\mu_{dc} - \mu_{uc} = - \mu_{Q_q} = \mu_e \ .
\label{beta}
\end{equation}
If we impose the requirements of electric and color charge neutrality,
$\mu_e$ and $\mu_8$ become fixed by the conditions of vanishing electric
and color densities:
\begin{eqnarray}
\rho_{Q_{tot}} &=& \rho_{Q_q}- \rho_e = \sum_{c=r,g,b}
\left(\frac23 \ \rho_{uc} - \frac13 \ \rho_{dc} \right) - \rho_e =
0 \ \ , \qquad \rho_8 = \frac{1}{\sqrt3} \sum_{f=u,d} \left(
\rho_{fr} + \rho_{fg} - 2 \rho_{fb} \right) = 0 \ , \label{cod}
\end{eqnarray}
where
\begin{equation}
\rho_e =  - \frac{\partial \Omega}{\partial \mu_e} =
- \frac{\partial \Omega^e}{\partial \mu_e}
\ \ , \qquad
\rho_{fc} = - \frac{\partial \Omega}{\partial \mu_{fc}}=
- \frac{\partial \Omega^\mf_{\rm (reg)}}{\partial \mu_{fc}} \ .
\end{equation}
Consequently, in the physical situation we are interested in, for each
value of $T$ and $\mu$ we should find the values of $\bar \Delta$, $\bar
\sigma$, $\mu_e$ and $\mu_8$ that solve Eqs.~(\ref{deltu}) and
(\ref{sigud}), supplemented by Eqs.~(\ref{beta}) and (\ref{cod}).

\hfill

The electron density can be evaluated analytically, leading to
\begin{equation}
\rho_e = \frac{\mu_e}{3 \pi^2} \left( \mu_e^2 + \pi^2 \ T^2 \right) \ .
\end{equation}
On the other hand, from Eqs.~(\ref{omegareg}), (\ref{once}) and
(\ref{freeomegareg}) the quark densities can be expressed as
\begin{equation}
\rho_{{\rm (reg)}fc}^\mf = \rho_{fc}^\mf - \rho_{fc}^{\rm free} +
{\rho_{\rm (reg)}^{\rm free}}_{fc} \ .
\end{equation}
Then a straightforward calculation leads to
\begin{eqnarray}
\rho_{fc}^{\mf} & = & 2\; T \sum_{n=-\infty}^{\infty} \int \frac{d^3
\vec{p}}{(2\pi)^3}\ \ \re \left(
\frac{1}{A_c}\;\frac{\partial A_c}{\partial \mu_{fc}} \right)\;,
\label{denu}
\end{eqnarray}
with
\begin{eqnarray}
\frac{\partial A_c}{\partial \mu_{uc}} & = & - \ 2\; (i\,\omega_n
+ \mu_{uc}) \;
 \left({p_{dc}^-}^2 + {\Sigma^*_{dc}}^2\right) \;
(1\; +\; 2\, \Sigma_{uc} \Sigma_{uc}')
\nonumber \\
& &
 -\ 2\; (1-\delta_{bc})\; \Delta^2 \bigg[\, 2\, \Sigma^*_{dc} \
\Sigma_{uc}'\; (i\,\omega_n + \mu_{uc})\;
+ \; i\,\omega_n - \mu_{dc}\, \bigg] \nonumber \\
& & - \ 4 i\ (1-\delta_{bc}) \ \bar\Delta \; \Delta \ \bigg[
\,{\Delta}\,^2 \,+\, (p_{uc}^+ \cdot p_{dc}^-) \, +
\, \Sigma_{uc} \,\Sigma_{dc}^*\, \bigg] \
\frac{\partial h(t, p^-_{dc} )}{\partial t_4}\Big|_{t=p_{uc}^+} \ ,
\nonumber \\
\end{eqnarray}
where we have defined $\Sigma_{fc}' = \bar \sigma \
\partial h(t,t)/\partial t^2|_{t^2={p_{fc}^+}^2}$.

The corresponding expressions for $\rho_{dc}$ are obtained by simply
exchanging $u$ and $d$ and taking the complex conjugate, while the
expressions for $\rho_{fc}^{\rm free}$ can be easily obtained from
$\rho_{fc}^\mf$ by setting $\bar \sigma = \bar \Delta =0$. Finally,
$\rho^{\rm free}_{{\rm (reg)}fc}$ is given by
\begin{equation}
\rho^{\rm free}_{{\rm (reg)}fc} = 2  \int \frac{d^3 \vec{p}}{(2\pi)^3}\;
\left\{
\left[ 1 + \exp \left( \frac{ \sqrt{\vec{p}\,^2+m^2}-\mu_{fc} }{T} \right) \right]^{-1}
- \ \left[1 + \exp \left( \frac{ \sqrt{\vec{p}\,^2+m^2}+\mu_{fc} }{T}
\right) \right]^{-1}\right\} \ .
\end{equation}

\section{Numerical results}

In this section we present our numerical results, showing the features of
the phase diagram and the behavior of relevant physical quantities for
Models I and II. According to previous analyses carried out within
nonlocal scenarios~\cite{DGS04}, the results are not expected to show a
strong qualitative dependence on the shape of the regulator. Thus we will
concentrate here on simple and well-behaved Gaussian regulator functions,
taking (in momentum space)
\begin{eqnarray}
r(p^2) & = & \exp(-p^2/2\Lambda^2) \qquad\qquad {\rm (Model \ I)} \\
\label{reg1}
g(p^2) & = & \exp(-p^2/\Lambda^2) \qquad\qquad {\rm (Model \ II)}
\label{reg2}
\end{eqnarray}
Here $\Lambda$ is a free model parameter, playing the r\^ole of an
ultraviolet cut-off. We have chosen a different normalization for Models I
and II in view of the relation between the respective regulating functions
[see Eq.~(\ref{hache})], which determine the
low $T$ and $\mu$ phenomenology.

\subsection{Parameterization}

For definiteness, for both Models I and II we choose here input parameters
$m$, $\Lambda$ and $G$ which allow to reproduce the empirical values for
the pion mass $m_\pi = 139$~MeV and decay constant $f_\pi = 92.4$~MeV, and
lead to a phenomenologically acceptable value for the chiral condensates
at vanishing $T$ and $\mu_{fc}$. For Gaussian regulators, taking into
account the chosen normalization of the cut-offs, it is seen that within
the MFA both models lead to the same expressions for the considered
physical quantities at $T=\mu=0$. However, this is not the case when one
goes beyond the MFA~\cite{GDS06}. In particular, the expressions for the
pion mass and decay constant are different (they are still coincident only
in the chiral limit), and it is necessary to use different sets of input
parameters. The parameters considered here for Model I are $m = 5.14$~MeV,
$\Lambda = 971$~MeV and $G \Lambda^2 = 15.41$, while for Model II we have
taken $m = 5.12$~MeV, $\Lambda = 827$~MeV and $G \Lambda^2= 18.78$. With
these sets we get for both models a phenomenologically reasonable value
for the chiral condensate, namely $\langle 0|\bar q q|0\rangle^{1/3} = -
250$~MeV. The remaining free parameter is the coupling strength $H$ in the
scalar diquark channel. In order to fix its value by hadron phenomenology
at zero $T$ and $\mu$, one would have to solve the Faddeev-type equations
for baryons as three-quark bound states which result from the quantization
of chiral quark models of the type considered in the present paper after
bosonization in meson and diquark channels (see~\cite{Cahill:1992ci} and
Refs.\ therein; for more elaborate recent calculations of nucleon
properties see~\cite{RWB03,Alkofer:2004yf}). Although in principle this is
possible, we refrain from fixing $H$ by hadron phenomenology within the
present exploratory study of the quark matter phase diagram and rather
choose different values for the coupling ratio $H/G$ in the range from 0.5
to 1.
\begin{figure}[hbt]
\begin{center}
\centerline{ \includegraphics[height=15truecm]{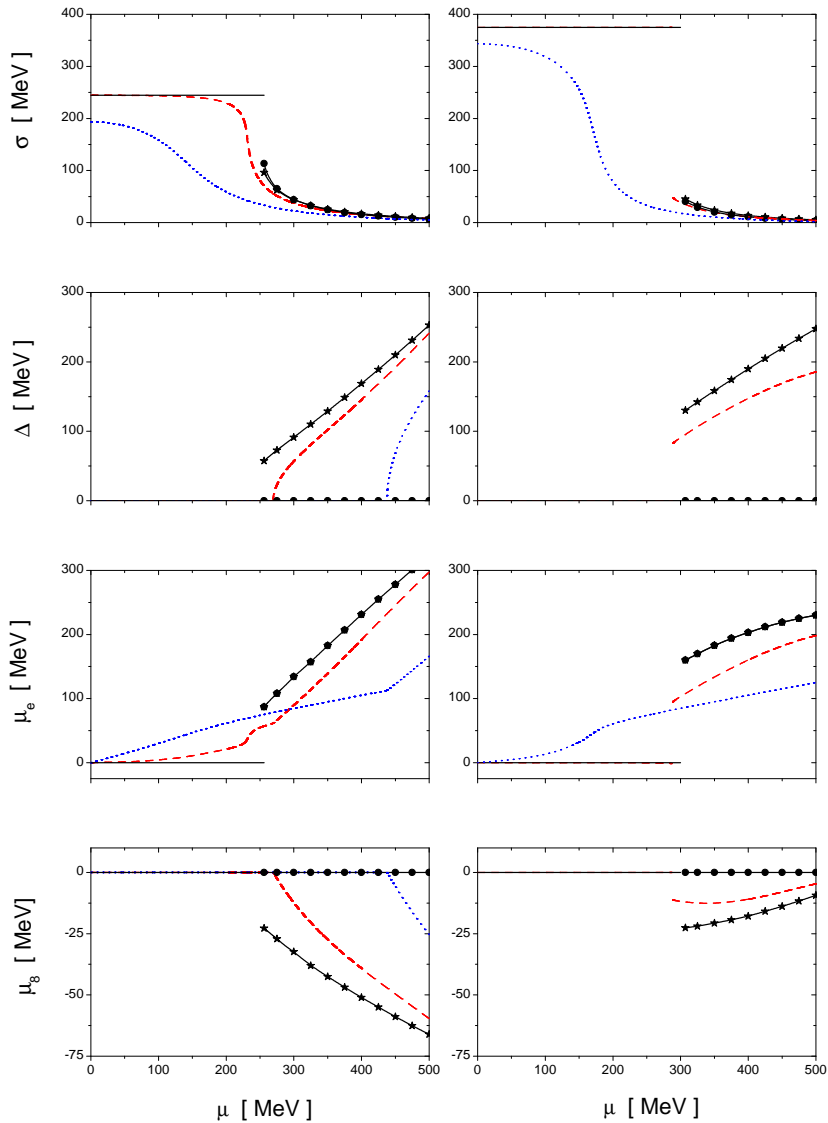} }
\caption{Behavior of the mean fields $\bar{\sigma}$ and $\bar{\Delta}$ and
the chemical potentials $\mu_e$ and $\mu_8$ for Models I (left) and II
(right) as a function of the chemical potential, for three different
values of the temperature. The curves correspond to the case $H/G=3/4$.
Full lines correspond to $T=0$, dashed lines to $T = 40$ MeV and dotted
lines to $T = 100$ MeV. In the case of $T=0$, lines marked with stars and
dots correspond to the 2SC and NQM phases respectively. }
\label{varios}
\end{center}
\end{figure}

\subsection{Order parameters and phase transitions}

For fixed values of the temperature $T$ and the chemical potential $\mu$
($\mu=\mu_B/3$), the mean field values $\bar \sigma$ and $\bar \Delta$, as
well as the chemical potentials $\mu_e$ and $\mu_8$, can be numerically
obtained from the gap equations (\ref{deltu}) and (\ref{sigud}), together
with the conditions of $\beta-$ equilibrium and vanishing color and electric
charge densities, Eqs.~(\ref{beta}) and (\ref{cod}). Let us begin by
considering the case $H/G = 0.75$, which is motivated by various effective
models of quark-quark interactions. Our results for $\bar\sigma$,
$\bar\Delta$, $\mu_e$ and $\mu_8$ are shown in Fig.~\ref{varios}, where we
plot these quantities as functions of $\mu$ for different representative
values of the temperature. Left and right panels correspond to Models I and
II, respectively.

For $T=0$ (solid lines), at low chemical potentials the system is for both
Models I and II in a chiral symmetry broken phase (CSB), where quarks acquire
large dynamical masses. By increasing the chemical potential one reaches a
first order phase transition, in which the chiral symmetry is approximately
restored, and a certain volume fraction of the quark matter undergoes a
transition to the 2SC phase coexisting with the remaining normal quark matter
(NQM) phase. The chemical potential $\mu_e$ (which for $T=0$ vanishes in the
CSB region) also shows a discontinuity across the transition. The new 2SC-NQM
mixed phase is a way in which the system realizes the constraint of electric
neutrality globally: the coexisting phases have opposite electric charges
which neutralize each other, at a common equilibrium pressure. In its
simplest realization, this mixed phase is treated within an approximation in
which Coulomb and surface energies are neglected (see
Ref.~\cite{Glendenning:1992vb}). For color superconducting quark matter this
realization of charge neutrality has been considered e.g.\ in
Ref.~\cite{Neumann:2002jm} for the NJL model and in Ref.~\cite{ABG05} for the
instantaneous nonlocal quark model. The discussion of inhomogeneous mixed
phases, {which are not yet fully understood}, crucially depends on the
assumptions for the surface tension and charge screening effects (see e.g.\
Ref.~\cite{Tatsumi:2002dq}) and goes beyond the scope of the present
investigation. On the other hand, following Refs.~\cite{Shovkovy:2003ce} and
\cite{Reddy:2005}, we have imposed color neutrality as a local constraint.
This is based on the fact that the color Debye screening length is expected
to be short and comparable to the inter-particle distance in the regime of
interest. As a consequence, $\mu_8$ turns out to be different in the two
components of the mixed phase. However, it should be kept in mind that this
chemical potential is in fact an effective quantity that has to be introduced
in these kind of models in order to account for the effect of the gauge
fields. Namely, as argued in
Refs.~\cite{Gerhold:2003,Kryjevski:2003,Dietrich:2004}, superconducting quark
matter is expected to be automatically color neutral in QCD. As expected, the
growth of the color chemical potential $\mu_8$ in the 2SC component of the
mixed phase is approximately proportional to that of the corresponding
$\bar\Delta$, which governs the amount of breakdown of the color symmetry due
to quark pairing.

When the temperature is increased (see dashed curves in Fig.~\ref{varios},
corresponding to $T=40$ MeV), the mixed phase is no longer favored and the
system goes into a pure 2SC phase. For $T=40$ MeV, this shows up as a
second order transition in the case of Model I, and a first order
transition in the case of Model II. Now, for both models, when one moves
along the first order transition line from $T=0$ towards higher
temperatures, one arrives at a triple point (3P). At this point the CSB
and 2SC phases coexist with a third NQM phase, in which the chiral
symmetry is approximately restored and there is no color
superconductivity. Finally, if $T$ is still increased, one reaches an
``end point'' (EP) where the first order transition from CSB to NQM phases
becomes a smooth crossover. The behavior of the dynamical masses and the
electric chemical potential $\mu_e$ along this smooth transition is shown
in Fig.~\ref{varios}, see curves corresponding to $T=40$ MeV (Model I) and
$T=100$ MeV (dotted lines, Models I and II).

\subsection{Quark matter phase diagrams}

\begin{figure}[hbt]
\begin{center}
\centerline{\includegraphics[height=20truecm]{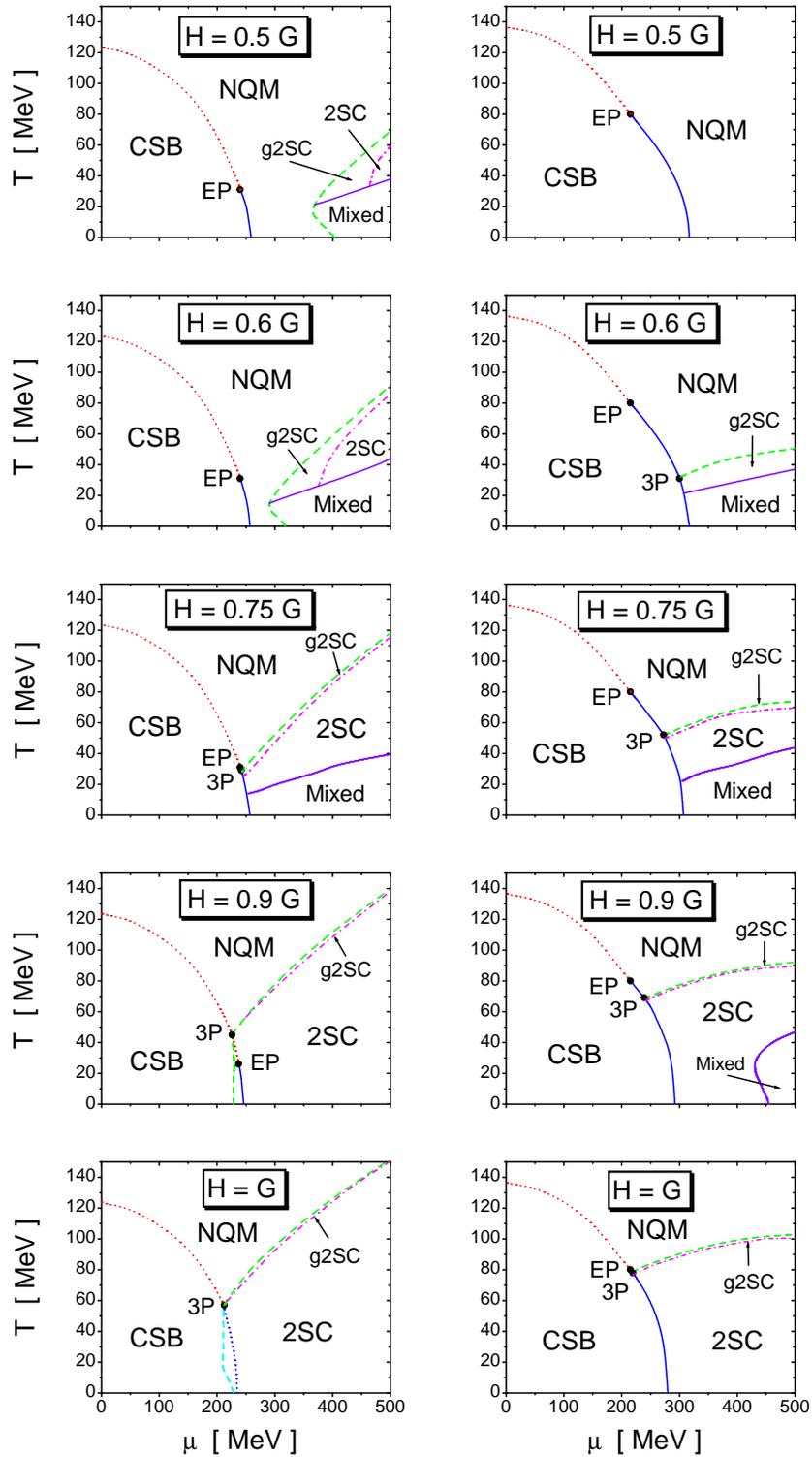} }
\caption{Phase
diagrams for Models I (left) and II (right) for different values of $H/G$.
Full and dashed lines indicate first and second order phase transition
curves respectively, dotted lines correspond to crossover-like
transitions, and dashed lines delimit the gapless 2SC band. Different
phases are denoted as NQM (normal quark matter phase), CSB (chiral
symmetry broken phase) and 2SC (two-flavor superconducting phase), while
the regions marked as ``Mixed" correspond to the NQM-2SC mixed phase. Note that in
Model I, for $H/G=1$ there is a narrow region in which one has 2SC while
the system is still in a chiral symmetry broken phase. EP and 3P denote
the end points and triple points respectively. }
\label{fases}
\end{center}
\end{figure}

The described features of the phase diagrams for Models I and II can be
visualized in the graphs shown in Fig.~\ref{fases}, where we plot the
transition curves on $T-\mu$ diagrams for different ratios $H/G$, and show
the regions corresponding to the different phases and the position of
triple and end points. In the graphs, solid and dotted lines correspond to
the mentioned first order and crossover transitions, respectively (in the
case of the crossover, the transition point can be defined by considering
the maximum of the chiral susceptibility~\cite{DS05}). Between NQM and 2SC
regions we find that in all cases there is a second order phase
transition, which corresponds to the dashed lines in the diagrams of
Fig.~\ref{fases}. Close to this phase border, the dashed-dotted lines in
the graphs delimit a band that corresponds to the so-called gapless 2SC
(g2SC) phase. Here, in addition to the two gapless modes corresponding to
the unpaired blue quarks, the presence of flavor asymmetric chemical
potentials $\mu_{dc}-\mu_{uc}\neq 0$ gives rise to another two gapless
fermionic quasiparticles~\cite{SH03}. Although the corresponding relations
cannot be derived analytically owing to the nonlocality of the
interactions, the border of the g2SC region can be numerically found. This
is done by determining whether for some value of $|\vec p|$ the imaginary
part of some of the poles of the Euclidean quark propagator vanish in the
complex $p_4$ plane. From the graphs it is seen that this g2SC band may
become relatively significant for low values of $H/G$. However, it never
extends up to zero temperatures, therefore this should not represent a
robust feature for compact star applications.

\section{Discussion}

Let us discuss some qualitative features of the curves displayed in
Figs.~\ref{varios} and~\ref{fases}. On one hand, for both models the 2SC
phase region becomes larger when the ratio $H/G$ is increased. This is not
surprising, since $H$ is the effective coupling governing the quark-quark
interaction that gives rise to the pairing. In any case, as a general
conclusion it can be stated that, provided the ratio $H/G$ is not too low,
these nonlocal schemes favor the existence of color superconducting phases
at low temperatures and moderate chemical potentials. Indeed, for the
parameters considered here, there is no 2SC phase only in the case of
Model II, $H/G=0.5$. This is in contrast with the situation in e.g.\ the
NJL model \cite{Bub04,Ru05}, where the existence of such a phase turns out
to be rather dependent on the input parameters. In addition, our results
are qualitatively different from those obtained in the case of the
noncovariant nonlocal models~\cite{ABG05}, where above the chiral phase
transition the NQM phase is preferable for values of the coupling ratio
$H/G\lesssim 0.75$. In those models, a color superconducting quark matter
phase can be found only for $H/G\approx 1$ .

\begin{figure}[htb]
\begin{center}
\centerline{ \includegraphics[height=4.5truecm]{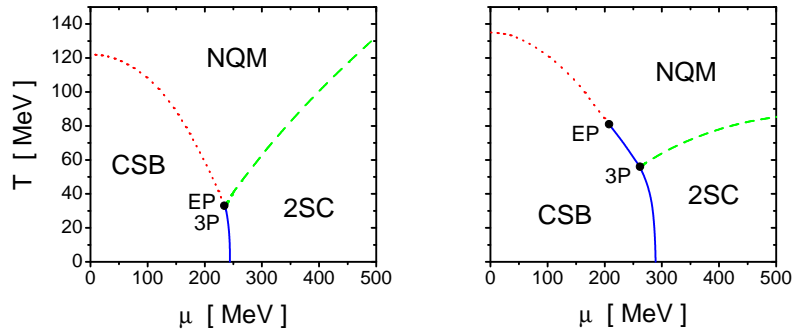} }
\caption{Phase diagrams for symmetric matter corresponding to Models I
(left) and II (right). Here we have taken $H/G=0.75$ }
\label{fases_2sc}
\end{center}
\end{figure}

It is also interesting to compare our results with those obtained for
isospin symmetric quark matter. For the same parameter sets, the
corresponding phase diagrams for $H/G=0.75$ are shown in
Fig.~\ref{fases_2sc}. By comparing them with those of Fig.~\ref{fases}, it
can be seen that the 2SC region becomes reduced when one imposes color and
electric charge neutrality conditions. This is indeed what one would
expect, since the condition of electric charge neutrality leads in general
to unequal $u$ and $d$ quark densities, disfavoring the $u$-$d$ pairing.
We notice, however, that the effect is relatively small, and the positions
of triple and end points as well as the shape of the critical lines remain
approximately unchanged. Concerning the shape of the chiral phase
transition line $T_{\rm CSB}(\mu)$, one observes at intermediate
temperatures \mbox{50~MeV$\;\lesssim\; T_{\rm CSB}\;\lesssim\;100$~MeV},
which are relevant for the future CBM experiment, an approximately linear
behavior. This can be seen as an interpolation between a convex shape
obtained in NJL or bag models and a concave shape for confining
Dyson-Schwinger equation models, see e.g.~Ref.~\cite{Blaschke:1997bj}. It
is remarkable that thus in the nonlocal covariant models presented here a
similarity with confining quark models occurs and that the
chiral/deconfinement transition line in the phase diagram resembles very
closely the positions of freeze-out parameters in heavy ion collisions.

Finally, we have studied the dependence of the phase transitions on the
model parameters, changing the input value of the chiral condensate within
a phenomenologically reasonable range 220 MeV $\leq - \langle 0|\bar q
q|0\rangle^{1/3} \leq$ 280 MeV. From this analysis, it is seen that the
qualitative features of the phase diagrams are not significantly modified.
In particular, it is seen that one finds in general color superconducting
phases at low temperatures and moderate chemical potentials, for
intermediate values of the ratio $H/G$. In addition, the values for the
critical temperature at $\mu=0$ are quite stable, yielding about 120 MeV
for Model I and 140 MeV for Model II. This would favor the description
given by Model II, in which the result is closer to the values provided by
lattice calculations.

\section{Conclusions}

We have considered the phase diagram of two-flavor quark matter under
neutron star constraints for two nonlocal, covariant quark models within
the mean field approximation. In the first one the nonlocality was due to
the regularization procedure, motivated by the instanton liquid model
(Model I), whereas in the second model a separable approximation of the
one-gluon exchange interaction was applied (Model II). Although for the
Gaussian regulators considered in this work the vacuum gap equations are
identical, both models differ in their fluctuation spectrum and therefore
in their parameters (current quark mass $m$, coupling strength $G$, UV
cutoff $\Lambda$), which have been fixed by using input values for the
pion mass, the pion decay constant and the chiral condensate. As result of
the numerical evaluation of the corresponding gap equations at finite
temperature and chemical potential, we have obtained that Model II
predicts a larger quark mass gap and a chiral symmetry breaking phase
transition line which extends 15-20\,\% further into the $T -\mu$ plane
when compared to Model I. The prediction for the critical temperature at
$\mu=0$ in Model II, $T_{\rm CSB}\sim$140 MeV, is closer to the results of
recent lattice QCD simulations than the prediction of both Model I and the
well-known local NJL model. Considering different values for the coupling
strength in the scalar diquark channel, we have found that under neutron
star constraints different low-temperature quark matter phases can occur
at intermediate densities: normal quark matter (NQM), pure superconducting
(2SC) quark matter and mixed 2SC-NQM phases. The critical temperature for
the 2SC phase transition is a rising function of $\mu$ in the case of
Model I whereas it is rather independent of $\mu$ for Model II, due to the
different $\mu$ dependences associated with the scalar diquark gaps. A
band of gapless 2SC (g2SC) appears at the border between 2SC and normal
quark matter. At large values of the $H/G$ ratio it is given by a tiny
strip in the phase diagram, and grows broader at low diquark couplings.
However, the g2SC region does not reach zero temperatures, thus it should
not represent a robust feature for compact star applications. Our present
investigation has been limited to the mean field approximation and the
neglect of the strange quark flavor. None of these approximations is
expected to be crucial for compact star applications, and the detailed
study of their impact on our results has been left for further development
of the present approach.

\section*{Acknowledgments}

The authors are glad to thank D.\ Aguilera for useful comments and
discussions. This work has been supported in part by CONICET and
ANPCyT (Argentina), under grants PIP 02368, PICT00-03-08580 and
PICT02-03-10718, and by a scientist exchange program between
Germany and Argentina funded jointly by DAAD under
grant No.\ DE/04/27956 and ANTORCHAS under grant
No.\ 4248-6.


\begin{thebibliography}{99}

\bibitem{Ritter:2004xj}
{\it Proceedings of the 17th International Conference on Ultra
Relativistic Nucleus-Nucleus Collisions (Quark Matter 2004), Oakland,
California, 11-17 Jan 2004}, J. Phys. {\bf G30}, S633 (2004)

\bibitem{Karsch:2003vd}
F.~Karsch, K.~Redlich and A.~Tawfik,
Eur.\ Phys.\ J.\ {\bf C29}, 549 (2003)

\bibitem{Shuryak:2004kh}
E.~Shuryak,
J.\ Phys.\ {\bf G30} S1221 (2004);
Prog.\ Part.\ Nucl.\ Phys.\ {\bf 53}, 273 (2004)

\bibitem{Blaschke:2003ut}
D.~B.~Blaschke and K.~A.~Bugaev,
Fizika {\bf B13}, 491 (2004);
Prog.\ Part.\ Nucl.\ Phys.\ {\bf 53}, 197 (2004)

\bibitem{Rev}
K.~Rajagopal and F.~Wilczek,
in {\it At the Frontier of Particle Physics. Handbook of QCD},
M. Shifman, ed. World Scientific, Singapore, 2001;
M. Alford, Annu. Rev. Nucl. Part. Sci. 51, 131 (2001);
D.~Rischke, Prog. Part. Nucl. Phys. {\bf 52}, 197 (2004)

\bibitem{Blaschke:2001uj}
D.~Blaschke, N.~K.~Glendenning and A.~Sedrakian (Eds.),
{\it Physics of neutron star interiors},
Lecture Notes in Physics {\bf 578}, Springer (2001);
D.~Blaschke, and D.~Sedrakian (Eds.),
{\it Superdense QCD matter and compact stars},
NATO Science Series {\bf II/197}, Springer (2006)

\bibitem{Klahn:2006ir}
T.~Kl\"ahn {\it et al.},
arXiv:nucl-th/0602038

\bibitem{Bailin:1983bm}
D.~Bailin and A.~Love,
Phys.\ Rept.\  {\bf 107}, 325 (1984)

\bibitem{ARW98}
M. Alford, K. Rajagopal and F. Wilczek,
Phys. Lett. {\bf B422}, 247 (1998);
R. Rapp, T. Sch\"{a}fer, E.V. Shuryak and M. Velkovsky,
Phys. Rev. Lett. {\bf 81}, 53 (1998)

\bibitem{Iida:2000ha}
  K.~Iida and G.~Baym,
  Phys.\ Rev.\ D {\bf 63}, 074018 (2001)
  [Erratum-ibid.\ D {\bf 66}, 059903 (2002)]

\bibitem{AR02}
M.~Alford and K.~Rajagopal,
JHEP {\bf 0206}, 031 (2002)



\bibitem{Gocke:2001ri}
C.~Gocke, D.~Blaschke, A.~Khalatyan and H.~Grigorian,
arXiv:hep-ph/0104183

\bibitem{Ru05}
S.~B.~R\"uster, V.~Werth, M.~Buballa, I.~A.~Shovkovy and D.~H.~Rischke,
Phys.\ Rev.\ {\bf D72}, 034004 (2005)

\bibitem{Bla05}
D.~Blaschke, S.~Fredriksson, H.~Grigorian, A.~M.~\"Oztas and F.~Sandin,
Phys.\ Rev.\ {\bf D72}, 065020 (2005)

\bibitem{Abuki:2005ms}
H.~Abuki and T.~Kunihiro,
Nucl.\ Phys.\ A {\bf 768}, 118  (2006)


\bibitem{Baldo:2002ju}
M.~Baldo, M.~Buballa, F.~Burgio, F.~Neumann, M.~Oertel and H.~J.~Schulze,
Phys.\ Lett.\ B {\bf 562}, 153 (2003)

\bibitem{Bub04}
M.~Buballa,
Phys.\ Rept.\  {\bf 407}, 205 (2005)

\bibitem{SH03}
I. Shovkovy and M. Huang,
Phys. Lett. \ {\bf B564}, 205 (2003)

\bibitem{Blaschke:2003yn}
D.~Blaschke, S.~Fredriksson, H.~Grigorian and A.~M.~\"Oztas,
Nucl.\ Phys.\ {\bf A736}, 203 (2004)

\bibitem{Grigorian:2003vi}
H.~Grigorian, D.~Blaschke and D.~N.~Aguilera,
Phys.\ Rev.\ {\bf C69}, 065802 (2004)

\bibitem{ABG05}
D.~N.~Aguilera, D.~Blaschke and H.~Grigorian,
Nucl.\ Phys.\ {\bf A757}, 527 (2005)

\bibitem{Bed02}
P.F.~Bedaque,
Nucl. Phys. {\bf A697}, 569 (2002)

\bibitem{Shovkovy:2003ce}
I.~Shovkovy, M.~Hanauske and M.~Huang,
Phys.\ Rev.\ D {\bf 67}, 103004 (2003)

\bibitem{Neumann:2002jm}
F.~Neumann, M.~Buballa and M.~Oertel,
Nucl.\ Phys. {\bf A714}, 481 (2003)

\bibitem{SS98}
T.\ Sch\"afer and E.V.\ Shuryak,
Rev.\ Mod.\ Phys.\ {\bf 70}, 323 (1998)

\bibitem{RW94}
C.~D.~Roberts and A.~G.~Williams,
Prog.\ Part.\ Nucl.\ Phys.\  {\bf 33}, 477 (1994);
C.~D.~Roberts and S.~M.~Schmidt,
Prog.\ Part.\ Nucl.\ Phys.\  {\bf 45}, S1 (2000)

\bibitem{SLW01}
J.~Skullerud, D.~B.~Leinweber and A.~G.~Williams,
Phys.\ Rev.\ D {\bf 64}, 074508 (2001)

\bibitem{NJL}
U.\ Vogl and W.\ Weise,
Prog.\ Part.\ Nucl.\ Phys.\ {\bf 27}, 195 (1991);
S.\ Klevansky,
Rev.\ Mod.\ Phys.\ {\bf 64}, 649 (1992);
T.\ Hatsuda and T.\ Kunihiro,
Phys.\ Rep.\ {\bf 247}, 221 (1994)

\bibitem{AS99}
E.\ Ruiz Arriola and L.L.\ Salcedo,
Phys.\ Lett.\ B {\bf 450}, 225 (1999)

\bibitem{Blaschke:1995gr}
D.~Blaschke, Y.~L.~Kalinovsky, G.~R\"opke, S.~M.~Schmidt and M.~K.~Volkov,
Phys.\ Rev.\ {\bf C53}, 2394 (1996)

\bibitem{Rip00}
G.\ Ripka,
Nucl.\ Phys.\ A {\bf 683}, 463 (2001);
R.S.\ Plant and M.C.\ Birse,
Nucl.\ Phys.\ A {\bf 703}, 717 (2002)

\bibitem{Ito:1991sz}
H.~Ito, W.~Buck and F.~Gross,
Phys.\ Rev.\ {\bf C43}, 2483 (1991);
Phys.\ Rev.\ {\bf C45}, 1918 (1992)

\bibitem{Buballa:1992sz}
M.~Buballa and S.~Krewald,
Phys.\ Lett.\ {\bf B294}, 19 (1992)

\bibitem{BB95}
R.D. Bowler and M.C. Birse,
Nucl. Phys. {\bf A582}, 655 (1995);
R.S.Plant and M.C. Birse,
Nucl. Phys. {\bf A628}, 607 (1998)

\bibitem{SGS04}
A.~Scarpettini, D.~Gomez Dumm and N.~N.~Scoccola,
Phys. Rev. D {\bf 69}, 114018 (2004)

\bibitem{BGR02}
W.~Broniowski, B.~Golli and G.~Ripka,
Nucl.\ Phys.\  {\bf A703}, 667 (2002)

\bibitem{RWB03}
A.H.~Rezaeian, N.R.~Walet and M.C.~Birse,
Phys. Rev. {\bf C70}, 065203 (2004)

\bibitem{RP05}
A.H. Rezaeian and H.-J. Pirner,
Nucl.\ Phys.\ A {\bf 769}, 35 (2006)

\bibitem{Blaschke:1999ab}
D.~Blaschke and P.~C.~Tandy,
in: ``Understanding Deconfinement in QCD'',
World Scientific, Singapore (2000), p. 218

\bibitem{GGS01}
I. General, D. G\'omez Dumm and N.N. Scoccola,
Phys. Lett. {\bf B506}, 267 (2001);
D. G\'omez Dumm and N.N. Scoccola,
Phys.Rev. {\bf D65}, 074021 (2002)

\bibitem{DS05}
D. G\'omez Dumm and N.N. Scoccola,
Phys. Rev. {\bf C72}, 014909 (2005)

\bibitem{DGS04}
R. S. Duhau, A. G. Grunfeld and N. N. Scoccola,
Phys. Rev. {\bf D70}, 074026 (2004)

\bibitem{Blaschke:2004cc}
D.~Blaschke, H.~Grigorian, A.~Khalatyan and D.~N.~Voskresensky,
Nucl.\ Phys.\ Proc.\ Suppl.\  {\bf 141}, 137 (2005)

\bibitem{Rip97}
G.\ Ripka,
{\it Quarks bound by chiral fields}
(Oxford University Press, Oxford, 1997)

\bibitem{SBK94}
S.~M.~Schmidt, D.~Blaschke and Y.~L.~Kalinovsky,
Phys.\ Rev.\ {\bf C50}, 435 (1994)

\bibitem{Amore:2001uf}
  P.~Amore, M.~C.~Birse, J.~A.~McGovern and N.~R.~Walet,
  Phys.\ Rev.\ D {\bf 65}, 074005 (2002)

\bibitem{BuSh05}
M.~Buballa and I.~A.~Shovkovy,
Phys.\ Rev.\ D {\bf 72}, 097501 (2005)

\bibitem{GDS06}
D. G\'omez Dumm, A.G. Grunfeld and N.N. Scoccola,
in preparation.

\bibitem{Cahill:1992ci}
R.~T.~Cahill,
Nucl.\ Phys.\ {\bf A543}, 63C (1992), and Refs. therein

\bibitem{Alkofer:2004yf}
R.~Alkofer, A.~Holl, M.~Kloker, A.~Krassnigg and C.~D.~Roberts,
Few Body Syst.\  {\bf 37}, 1 (2005)

\bibitem{Glendenning:1992vb}
N.~K.~Glendenning,
Phys.\ Rev.\ D {\bf 46},  1274 (1992)

\bibitem{Tatsumi:2002dq}
T.~Tatsumi, M.~Yasuhira and D.~N.~Voskresensky,
Nucl.\ Phys.\ A {\bf 718}, 359 (2003)

\bibitem{Reddy:2005}
S.~Reddy and G.~Rupak,~Phys.~Rev.~C {\bf 71}, 025201  (2005)

\bibitem{Gerhold:2003}
A.~Gerhold and A.~Rebhan,~Phys.~Rev.~D {\bf 68},  011502 (2003)

\bibitem{Kryjevski:2003}
A.~Kryjevski,~Phys.~Rev.~D {\bf 68}, 074008  (2003)

\bibitem{Dietrich:2004}
D.~D.~Dietrich and D.~H.~Rischke,~Prog.~Part.~Nucl.~Phys.~{\bf 53}, 305 (2004)

\bibitem{Blaschke:1997bj}
D.~Blaschke, C.~D.~Roberts and S.~M.~Schmidt,
Phys.\ Lett.\ B {\bf 425}, 232 (1998)

\end{thebibliography}
\end{document}